\documentclass[galaxies,commentary,submit,oneauthor,LaTex, dvi2pdf,10pt,a4paper]{mdpi} 

\firstpage{1} 
\articlenumber{x}
\doinum{10.3390/------}
\pubvolume{xx}
\pubyear{2018}
\copyrightyear{2018}
\externaleditor{Academic Editor: Dmitry Malyshev}
\history{Received: date; Accepted: date; Published: date}
\pdfoutput=1

\usepackage{aas_macros}
\usepackage{amsmath}
\preto{\abstractkeywords}{\nolinenumbers}
 \theoremstyle{mdpi}
 \newcounter{thm}
 \newcounter{ex}
 \newcounter{re}
 \theoremstyle{mdpidefinition}

\Title{Positron Transport And Annihilation In The Galactic Bulge}

\Author{Fiona H. Panther $^{1}$}
\AuthorNames{Fiona H. Panther}

\address[1]{%
Research School of Astronomy and Astrophysics, Australian National University, Canberra 2611, Australia; fiona.panther@anu.edu.au}

\corres{Correspondence: fiona.panther@anu.edu.au;}

\abstract{The annihilation of positrons in the Milky Way galaxy has been observed for $\sim 50$ years however the production sites of these positrons remains hard to identify. The observed morphology of positron annihilation gamma-rays provides information on the annihilation sites of these Galactic positrons. It is understood that the positrons  responsible for the annihilation signal originate at MeV energies. The majority of sources of MeV positrons occupy the thin, star forming disk of the Milky Way. If positrons propagate far from their sources, we must develop accurate models of positron propagation through all ISM phases in order to reveal the currently uncertain origin of these Galactic positrons. On the other hand, if positrons annihilate close to their sources, an alternative source of MeV positrons with a distribution that matches the annihilation morphology must be identified. In this work, I discuss the various models that have been developed to understand the origin of the $511\,\mathrm{keV}$ line from the direction of the Galactic bulge, and the propagation of positrons in the ISM.}


\begin{document}





\section{Introduction}
The annihilation of positrons (anti-electrons) at rest results in the characteristic emission of a gamma-ray line at $\sim 0.5\,\mathrm{MeV}$. This gamma-ray line was the first originating beyond the solar system to be detected from Earth in the early 1970s \citep{Johnson72}. Since then, the spatial extent and spectral characteristics of gamma-ray emission at $\sim 0.5\,\mathrm{MeV}$ has been explored by both balloon-borne and space missions. The most recent observations made by \textit{INTEGRAL}/SPI reveal the annihilation of positrons at a rate of $2\times10^{43}\,\mathrm{s^{-1}}$ in the central $\sim 20\deg$ of the Galaxy based on the effective distance between Earth and the Galactic bulge being $8.5\,\mathrm{kpc}$ \citep{Siegert16} (the Galactic bulge annihilation line). In \citep{Siegert16}, 11 years of \textit{INTEGRAL}/SPI exposures are combined, and the resulting positron annihilation rate is consistent with  earlier analyses \citep{Knoedelseder05,Weidenspointner08} which also reveal the annihilation of positrons in the bulge region of the Galaxy. The Galactic disk has been detected for some time (e.g. \cite{Knoedelseder05,Weidenspointner08,Churazov11} however emission from the Galactic disk was thought to be concentrated in the inner 100 degrees of the Galaxy \cite{Skinner2010, Bouchet2011}. The most recent analysis \cite{Siegert16} suggests the emission of positron annihilation gamma-rays across the full extent of the Galactic disk and a higher rate of positron annihilation in the Galactic disk than previous analysis revealed. \\
Analysis of the gamma-ray spectrum produced by the annihilating positrons indicates they annihilate predominantly in the warm, partially ionized phase of the interstellar medium (ISM) \citep{Churazov05,Jean06,Churazov11,Siegert16} by undergoing interactions with neutral hydrogen after they have slowed to kinetic energies of $\sim 10-100\,\mathrm{eV}$. The reaction rate between positrons and neutral hydrogen dominates in this phase of the ISM due to the large cross-section for interaction between positrons and hydrogen at energies of $\sim 10\,\mathrm{eV}$ and the low abundances of other neutral species such as helium. Current constraints on the energy at which these positrons are injected into the ISM come from fundamental physics \citep{Aharonian81} and data obtained by the Compton Gamma-Ray Observatory (\textit{CGRO}) and \textit{INTEGRAL} missions, and suggest the initial kinetic energy of the annihilating positrons does not exceed a few MeV. This rules out the origin of the positrons being cosmic ray (CR) secondaries produced in the decay of pions, as well as other sources of relativistic positrons, such as pulsars, millisecond pulsars and magnetars \citep{Prantzos11}. However, the combination of data from both \textit{INTEGRAL} and \textit{CGRO} may result in systematics which will be present in the data used to derive the constraint, and moreover \citep{Beacom06} assumes that the electrons with which positrons annihilate at high ($>MeV$) energies are at rest. If positron annihilation is occuring in an environment where relativistic electrons are present (e.g. in a microquasar jet \citep{Guessoum06,Siegertmicroquasars} or pulsars and millisecond pulsars, \citep{Bartels18}) it may be possible to circumvent this constraint. However, the presence of the $511\,\mathrm{keV}$ line and ortho-positronium continuum at low gamma-ray energies strongly suggest that the majority of positrons annihilate via interactions with neutral hydrogen at thermal energies.\\
The origin of these annihilating positrons has puzzled astronomers for half a century. The morphology - gamma-ray emission dominated by a bulge component - does not reflect emission at any other wavelength.\\
A viable source of MeV positrons of astrophysical origin is the decay of $\beta^+$-unstable radionuclides,  such as $^{26}$Al, $^{44}$Ti and $^{56}$Ni, produced by massive stars and supernovae. Massive stars and supernovae tend to occur in the star-forming thin disk of the Milky Way \citep{Prantzos11} while the highest surface brightness of positron annihilation gamma-rays is observed in the Galactic bulge region. To explain the mismatch between the distribution of prospective positron birth sites, and the observed annihilation sites, we must investigate the propagation of positrons. In this short review, I summarize the various transport mechanisms that have been explored in the literature to date to explain the annihilation of Galactic bulge positrons. 

\section{Positron transport in the ISM}
From the annihilation spectrum observed in the inner Galaxy, analysis of the gamma-ray spectrum implies that while positrons are born into the ISM with $\sim\,\mathrm{MeV}$ energies, they must cool to energies $< 10\,\mathrm{eV}$ before annihilating. Positrons lose energy via ionization and Coulomb interactions, synchrotron losses due to interactions with the magnetic field in the Galaxy, inverse Compton scattering and bremsstrahlung. Positrons may also lose energy through adiabatic energy losses (e.g. \citep{Panther2017}). Positrons with energies $\leq \mathrm{GeV}$ lose energy predominantly through ionization and Coulomb interactions with neutral atoms and free electrons in the ISM, whereas positrons with energies $>\mathrm{GeV}$ lose their kinetic energy via inverse Compton, synchrotron and bremsstrahlung losses.\\
As they decelerate, positrons are thought to propagate away from their sources. This propagation may occur through either collisionless or collisional transport. In collisionless transport, positrons scatter off magnetic turbulence in the ISM, wheras in collisional transport scattering occurs between positrons and individual particles in the ISM (e.g. neutral hydrogen atoms and electrons), moderated by Coulomb interactions. Both collisional and collisionless transport have been discussed in the literature, and a summary is provided in the subsequent sections of this work.
\section{Outside-In Transport}
If MeV positrons are produced by processes associated with star formation, the majority of positrons will be born in the star forming thin disk of the Milky Way. One way to then explain the high bulge/disk annihilation ratio is that some disk positrons find their way into the Galactic bulge before they annihilate. Such a mechanism is positron interactions with the large scale magnetic field structure of the Milky Way. In general, such a scenario requires positrons to interact with the poloidal halo component that dominates the inner Galaxy as described in \cite{Prantzos06}. The turbulent magnetic field with which positrons interact in the collisionless transport process has a field strength roughly comparable to that of the large scale magnetic fields ($1-10\,\mu \mathrm{G}$) in the disk region where the positrons are produced, and this may tend to confine MeV positrons to the Galactic disk in a similar manner to $\sim$ GeV energy cosmic ray protons that make up the cosmic ray halo \cite{Prantzos06}. However, if positrons can escape the cosmic ray halo, they will begin to interact with the poloidal magnetic field and be transported into the Galactic bulge \cite{Prantzos06}. This idea was explored in detail by Higdon et al. \cite{Higdon09}, who include a detailed description of the distribution  and filling factors of various ISM phases, as well as accounting for the distribution of positron sources. Specifically, they find that around half of positrons from the sources considered therein are produced in the inner $3\,\mathrm{kpc}$ of the Galaxy (the inner disk and Galactic bulge), and that these positrons tend to annihilate in the dense, warm outer shells of molecular clouds that occupy the inner $\sim 1.5\,\mathrm{kpc}$ of the Galaxy. The remaining half of the total Galactic positron yield calculated in \cite{Higdon09} is produced in the outer disk of the Galaxy. Around half of these disk positrons annihlate in the disk itself, while the remainder escape into the Galactic halo, lowering the annihilation flux from the disk and consequently the observed $B:D$ annihilation flux ratios. One of the key details of this work is the description of how positrons propagate through the different ISM phases. Positrons in the hot, ionized phases of the ISM tending to diffuse along magnetic flux tubes where the diffusion is controlled by resonant scattering by magnetohydrodynamical (MHD) waves. On the other hand, in the neutral or mostly neutral phases, these waves are damped and positrons tend to stream along the flux tubes with an isotropic pitch angle distribution. The distance positrons can propagate in each phase is controlled by either the diffusion mean free path (which is a function of the magnetic field and incident particle energy) in the ionized phase, or the streaming velocity. The annihilation site of the positrons is determined from comparison of the above propagation distances to the typical scale of the phase in which the positron is born. If the propagation distance exceeds the size scale of the ISM phase in which the positron is born, it escapes. Otherwise, the positron is assumed to annihilate in-situ.\\
To evaluate whether a given positron production and transport scenario can explain the observed positron annihilation signal from the Galaxy, both the total positron production rate must be replicated, as well as the morphology. As the results based on SPI/\textit{INTEGRAL} data are presented as varying best-fit models, one way of quantifying the morphology is to use the ratio of positron annihilation fluxes in the bulge and disk. This is commonly referred to in the literature as the bulge to disk ratio, or B/D. In the early years of SPI/\textit{INTEGRAL}, B/D was determined to  be in excess of 1 ($B/D\sim 1.4$, \citep{Knoedelseder05,Weidenspointner08}).\\
The authors of \cite{Higdon09} find that the scenario they describe, where positrons propagate differently in the ISM of the Galactic bulge compare to the disk, can (within large uncertainties) replicate high B/D values observed by SPI/\textit{INTEGRAL}. However, more recent simulations of positron transport have involved either detailed Monte Carlo simulation of positron trajectories in the ISM \cite{Alexis14}, where the probabilities for particle scattering and annihilation take into account both collisional and collisionless scattering, and various annihilation processes (see the Appendix B of \cite{Jean09} for more details). Alternatively cosmic ray propagation codes such as GALPROP \cite{Martin2012} have been adapted to model positron propagation in the Galaxy. In both cases, it is found that replicating high B/D values is challenging, even for extreme prescriptions that allow transport of positrons several kpc from their birth sites. 
\section{Galactic Center positron sources}
Exploration of the idea that positrons annihilating in the Galactic bulge originate in the central regions of the Galaxy predates the \textit{INTEGRAL} mission. In the mid-1990's, the first image reconstructions of the extent of positron annihilation radiation were made with data from the OSSE instrument on \textit{CGRO}, and seemed to reveal an extended emission component stretching up to $8\,\mathrm{kpc}$ above the Galactic plane (dubbed the Positive Latitude Enhancement or PLE) \cite{Purcell97}. An explanation for this "annihilation fountain" phenomenon was developed in \cite{Dermer97}. This work described a nuclear outflow associated with the Milky Way, capable of sending positrons produced in the Galactic center several kpc above the Galactic plane. Remarkably, this work predates even the first evidence for a nuclear outflow in the Milky Way \cite{BHC03} and predates the discovery of the Fermi Bubbles \cite{Su10}, another phenomenon associated with the Galactic center region, by almost 15 years. The PLE phenomenon was later shown, however, to be an artefact of the image reconstruction analysis, and not an astrophysical signal \cite{Milne01b}. To date, very little is known about positron annihilation at high Galactic latitudes as the \textit{INTEGRAL} mission has concentrated efforts along the Galactic plane (e.g. see recent exposure maps in \citep{Siegert16}). Future work to extend \textit{INTEGRAL} exposures to high latitudes to search for extended $511\,\mathrm{keV}$ emission is now a priority.\\
While the PLE was shown to be an artefact, a Galactic outflow may still disperse positrons into the Galactic bulge. The enhanced star formation in the central molecular zone (CMZ) of the Galactic center, as well as high energy processes associated with the central SMBH, make the Galactic center a potentially interesting positron source itself. Explaining how positrons produced in the inner few hundred parsecs are transported outward to fill the Galactic bulge where they are observed to annihilate presents a challenge. The possibility that positrons diffuse away from the central regions of the Galaxy into the Galactic bulge was investigated in detail using a Monte Carlo simulation which tracks positron trajectories through the ISM in \cite{Jean09}. While in \cite{Higdon09}, the interaction between positrons and MHD waves was invoked to transport positrons over large distances, \cite{Jean09} found that the damping of these waves results in positron propagation being controlled by collisions with gas particles. In the collisional transport mode described in the paper, it is found that positrons travel along magnetic field lines and annihilate far from their sources, and the distribution of annihilation positrons depends on the spatial distribution of the Galactic magnetic field. Thus, if the magnetic field is poloidal, positrons tend to stay confined to a cone subtending a small solid angle above and below the Galactic center rather than filling the entire Galactic bulge.\\
The diffusion lengths initially calculated in \citep{Jean09} are far in excess of the typical size scales of the warm ISM, where positrons are thought to annihilate. However, positrons tend to be born in turbulent environments, and the propagation of positrons is likely to be dominated by scattering from MHD waves at higher energies (with propagation lengths of $\sim 20 - 80\,\mathrm{pc}$ for positrons with energies $\sim \mathrm{MeV}$). As positrons lose energy to the plasma and their kinetic energies drop below a threshold for the Larmor radius to be of a similar scale to the turbulence, positrons then will tend to propagate through collisional transport. The total propoagation distances of MeV positrons, combining the time the positron spends in the collisional and collisionless regime, is evaluated to be $\sim 40-160\,\mathrm{pc}$. Thus, these authors ultimately conclude that positrons may only travel a few hundred parsecs from their birth sites.\\
The coupling of positrons to the turbulent plasma in which they were born makes it possible that the propagation of positrons can be dominated by the motion of the plasma itself. While the concept of the "annihilation fountain" \cite{Dermer97} was furloughed, powerful evidence for a nuclear outflow associated with the Milky Way has since emerged, first in the mid-infra red \cite{BHC03}, and later with the discovery of the gamma-ray emitting `Fermi Bubbles" \cite{Su10}. The possibility that positrons could be advected into the Galactic bulge by a nuclear outflow was suggested in \cite{Jean09,Churazov11}, as this scenario could potentially explain not only the morphology by transporting positrons out to $\sim$kpc scales but also the spectrum of positron annihilation gamma-rays: Positrons born into the hot, $\sim10^7\,\mathrm{K}$ plasma at the base of the outflow may not annihilate until the outflowing plasma has cooled to $\sim 10^4\,\mathrm{K}$ through both adiabatic and radiative cooling processes. An explicit connection between the nuclear outflow, the Fermi Bubbles and Galactic bulge positrons was mentioned in \cite{Crocker2011}, and investigated in detail in \cite{Panther2017}. The authors perform a parameter study to investigate whether any steady state nuclear outflow with a given input mass and energy flux from star formation can advect positrons to radii consistent with \textit{INTEGRAL} observations. While advection of MeV positrons to size scales of $\sim 2\,\mathrm{kpc}$ is possible, the development of a strong ionization and temperature gradient in the steady-state outflow ultimately results in the annihilation of the majority of positrons in the hot, ionized phase of the ISM. Constraints from the observed annihilation spectrum indicate no more than a few per cent of positrons annihilate in the hot phase, in the case of annihilation in a multi-phase ISM \cite{Churazov05}.\\
In \citep{Panther2017}, the study was restricted to a steady-state nuclear outflow and the limits of the work can thus put weak constraints on positron injections in a single burst. Single-burst injection of positrons was investigated in more detail in \cite{Cheng06,Totani06} and was commented on in particular by Alexis et al. \cite{Alexis14}, who find that a burst of either star formation, or activity of the Milky Way's central supermassive black hole (SMBH) producing $10^{57} -10^{60}$ positrons with energies $<1\,\mathrm{MeV}$ could contribute to the Galactic bulge annihilation signal. However, this scenario requires a coincidence with a previous burst of star formation or AGN-like activity from the SMBH, which must have occured $0.3 - 10\,\mathrm{Myr}$ ago.
\section{Distributed positron sources}
The conclusion that, according to detailed simulations, positrons annihilate close to their sources and that large scale gas motions may be unable to consistently account for all properties of positron annihilation in the Galactic bulge (both the morphology and spectrum of the emission) is not the only motivation to search for a positron source that has a similar distribution to that of the annihilation signal. With 11 years of \textit{INTEGRAL} data, a new best fit model for the morphology of positron annihilation in the Galaxy has emerged \cite{Siegert16}: observations now favor a model that includes an extended thick disk of emission with a spatial extent similar to that of the Galactic thick disk. Moreover, updated estimates on the flux of not only this thick disk and the Galactic bulge component, but also of a source coincident with the Galactic center reveal flux ratios between bulge and disk, and bulge and "Galactic Center Source" that are consistent with the stellar mass ratios between these regions. In light of the new observations and recent theoretical work on positron propagation, the search has now turned to finding a Galactic positron source associated with the old stellar population. Once again, the usage of B/D to evaluate good morphological fits has been used. In \citep{Siegert16}, $B/D\sim 0.4$, a large reduction from earlier values \citep{Knoedelseder05,Weidenspointner08}. This is due to the introduction of the extended thick disk model component. It should be emphasised that this value is taken from empirical best-fit models to simplify subsequent analysis and should not be over-interpreted.\\
The remarkable observation of positron annihilation in the 2015 outburst of microquasar V404 Cygni \cite{Siegertmicroquasars} lends weight to the conjecture that such a source of positrons is associated with microquasars and low-mass x-ray binaries (LMXRBs) during an outburst, or producing jets \cite{Guessoum06}. During its 2015 outburst, positron annihilation in the microquasar jet was invoked to explain the broad emission feature at $\sim 0.5\,\mathrm{MeV}$ in the gamma-ray spectrum, which extends above $511\,\mathrm{keV}$, and can naturally be explained by pair production in the jet which then results in pair-plasma annihilation \cite{Aharonian81,Svensson82}, annihilation in-flight \cite{Beacom06} or pair cascades \citep{Aharonian81}. To explain the positron annihilation rate and morphology, around 10 such sources would be required to be active at any time, with a total of $10^3 - 10^4$ systems being present in the Milky Way \citep{Siegertmicroquasars}. This is consistent with binary population synthesis estimates for the number of these systems \cite{Sadowski08}. However, this is dependent on the total positron yield of each event, a value which is highly uncertain.\\
An alternative scenario invokes a sub-type of SNe Ia that are observationally connected to old stellar populations. Around 30 per cent of all SNe Ia in early-type host galaxies are observed to be of the sub-type SN1991bg-like (SNe 91bg). These supernovae are both photometrically and spectroscopically peculiar, with their red colors and deep Ti\textsc{II} spectral absorption feature. While positrons produced in the decay of the 56-Ni daughter nucleus 56-Co will be predominantly annihilated in the SN ejecta, as with normal SNe Ia \cite{Milne99, Taubenberger08}, positrons from the much longer-lived $^{44}$Ti decay chain will escape the SN into the surrounding ISM. In \cite{Crocker17}, binary population synthesis calculations are performed to identify a candidate progenitor channel for these events. Low mass ($\sim 1.4-2\,\mathrm{M_\odot}$) binary systems evolve into systems composed of a carbon-oxygen white dwarf ($M\sim 0.9\,\mathrm{M_\odot}$) and a pure helium white dwarf ($M\sim 0.31 - 0.37\,\mathrm{M_\odot}$) which merge at long characteristic delay times (3-6 Gyr). The supernova resulting from this merger would be required to produce $\sim 0.031\,\mathrm{M_\odot}$ 44-Ti per transient event (equivalently $\sim 5.8 \times 10^{-5}\,\mathrm{M_\odot/yr}$) to explain the origin of the majority of positrons in the Milky Way, and this yield is roughly consistent with existing calculations of Ti-44 yield from sub-Chandrasekhar helium detonations.  This long delay time implies a connection between old stars in the Milky Way and these thermonuclear supernovae, and the rate derived for these events from the binary population synthesis calculation is consistent with the expected Galactic rate of SNe 91bg, derived from observations of these supernovae in external Galaxies.\\
Combining the total positron yield expected from this transient and the expected rate in the disk, bulge and nuclear region of the Galaxy, the authors of \citep{Crocker17} find that they can account for 90\% of the total positron production in the Galaxy (the remaining 10\% being accounted for by 26-Al synthesis in massive stars) and replicate B/D and N/B within uncertainties. However, detailed three-dimensional hydrodynamical simulations of the proposed supernova explosion and further post-processing of these simulations with nuclear reaction networks will be required to further evaluate the plausibility of this scenario. Furthermore, these simulations must also be post-processed with radiative transfer networks to investigate whether the proposed scenario provides a match to observed SNe 91bg spectra and light curves. 
\section{Conclusions}
Understanding the transport of MeV positrons in the ISM is required to reveal the source of the $\sim 5 \times 10^{43}$ positrons that annihilate each second in the Galaxy. The bulge-dominated morphology of this astrophysical signal is contrasted with the distribution of MeV positron sources, which are usually associated with the star-forming thin disk of the Galaxy, not the more quiescent stellar bulge. The propagation of positrons has been investigated through both theory and simulation, with recent conclusions suggesting that positrons remain confined to within a few hundred parsecs of their birth sites by interstellar turbulence and that the distribution of positron sources is directly reflected by the morphology of annihilation gamma-rays. Such sources are thought to be associated with the old stellar population of the Milky Way. However, uncertainty remains about the dominant modes of positron propagation and the details of collisionless and collisional transport of positrons, which is complicated by the need for more detailed understanding of interstellar magnetic turbulence and the Galactic magnetic field that affects positron propagation. Consequently, there is still much work to be done to understand the birth, life and death of positrons in the interstellar medium.
\vspace{6pt}

\acknowledgments{This research is supported by an Australian Government Research Training Program (RTP) Scholarship. FHP thanks Roland Crocker, Roland Diehl, Eugene Churazov, Thomas Siegert and Torsten Enßlin for useful discussions, and Christoph Weniger and Dmitry Malyshev, organisers of the conference "Three Elephants in the Gamma-Ray Sky". FHP also thanks the reviewers for their detailed feedback on ways to improve the manuscript.}



\bibliographystyle{mdpi}


\bibliography{lite}

\end{document}